\documentclass[12pt,preprint]{aastex}
\begin{document}

\title{Novae With Long-Lasting Supersoft Emission That Drive a High Accretion Rate}
\author{Bradley E. Schaefer  \& Andrew C. Collazzi\affil{Physics and Astronomy, Louisiana State University, Baton Rouge, LA 70803}}

\begin{abstract}

We identify a new class of novae characterized by the post-eruption quiescent light curve being more than roughly a factor of ten brighter than the pre-eruption light curve.  Eight novae (V723 Cas, V1500 Cyg, V1974 Cyg, GQ Mus, CP Pup, T Pyx, V4633 Sgr, and RW UMi) are separated out as being significantly distinct from other novae.  This group shares a suite of uncommon properties, characterized by the post-eruption magnitude being much brighter than before eruption, short orbital periods, long-lasting supersoft emission following the eruption, a highly magnetized white dwarf, and secular declines during the post-eruption quiescence.  We present a basic physical picture which shows why all five uncommon properties are causally connected.  In general, novae show supersoft emission due to hydrogen burning on the white dwarf in the final portion of the eruption, and this hydrogen burning will be long-lasting if new hydrogen is poured onto the surface at a sufficient rate.  Most novae do not have adequate accretion for continuous hydrogen burning, but some can achieve this if the companion star is nearby (with short orbital period) and a magnetic field channels the matter onto a small area on the white dwarf so as to produce a locally high accretion rate.  The resultant supersoft flux irradiates the companion star and drives a higher accretion rate (with a brighter post-eruption phase), which serves to keep the hydrogen burning and the supersoft flux going.  The feedback loop cannot be perfectly self-sustaining, so the supersoft flux will decline over time, forcing a decline in the accretion rate and the system brightness.  We name this new group after the prototype, V1500 Cyg.  V1500 Cyg stars are definitely not progenitors of Type Ia supernovae.  The V1500 Cyg stars have similar physical mechanisms and appearances as predicted for nova by the hibernation model, but with this group accounting for only 14\% of novae.

\end{abstract}

\section{Introduction}

Novae occur in systems with constrained properties (Roche lobe overflow onto a white dwarf, where the matter accumulates until a thermonuclear runaway is triggered), so in overview, they are a fairly homogenous group.  Nevertheless, variations in the system properties (e.g., orbital period, strength of the white dwarf magnetic field) are substantial, and this leads to diverse behavior.  For example, the uncommon novae with long orbital period can lead to systems that suffer dwarf nova eruptions (like GK Per and V1017 Sgr) or have a short recurrence time scale (like T CrB and RS Oph).  A high magnetic field will impose an accretion column to the white dwarf with no disk (for polars like V1500 Cyg and GQ Mus) or a partial disk feeding an accretion column (for intermediate polars like CP Pup).  These groups or subsets of novae have their properties caused by particular system parameters.  The state of the field is now that our community is trying to work out the groups and their physical mechanisms.

An important property of novae is that X-ray satellites (and especially the {\it Swift} satellite) are finding that eruptions produce a high luminosity of supersoft X-ray photons (with temperatures of typically 30 eV).  In general, novae eruptions should all become supersoft sources (SSSs) for some time near the end of each eruption after the photosphere has receded to the point where the underlying hydrogen burning on the surface of the white dwarf is revealed (MacDonald et al. 1985; Hachisu \& Kato 2006).  But the first two observed supersoft novae, GQ Mus (\"{O}gelman et al. 1993) and V1974 Cyg (Krautter et al. 1996), were bright SSSs long after the end of their eruptions, and more long-lasting supersoft novae have been identified.  The turn-off time for the SSS phase varies widely (Greiner et al. 2003; Ness et al. 2007).  The existence of the long-duration SSSs has already been tied to systems with short orbital periods (Greiner et al. 2003).

An important physical mechanism is the irradiation of the companion star by the eruption itself and by the white dwarf soon after the eruption.  The idea is that this irradiation will drive relatively high mass-loss and the resulting high accretion rate will keep the white dwarf luminous enough to drive yet more mass-loss.  The source of the white dwarf's luminosity is usually invoked as simply a hot white dwarf (Shara et al. 1986; Schmidt et al. 1995; Thomas et al. 2008), but other possibilities are reasonable (like SSS emission).  The mechanism to drive the higher mass-loss rate could be the increase of the companion star radius (King et al. 1995; 1996; 1997; Kolb et al. 2001) or the creation of a wind from the upper atmosphere (van Teesling \& King 1998; Knigge et al. 2000).  The theory behind this irradiation driven mass-loss is complex, with difficulties being the calculation of the exact structure of the companion star's atmosphere along with the vertical and horizontal transport of energy and the effect on the mass-loss for all regimes of system parameters.  Observationally, the effects of irradiation have already been measured to be large for V1500 Cyg (Schmidt et al. 1995), GQ Mus (Diaz et al. 1995), and T Pyx (Knigge et al. 2000).

An important process for novae is hibernation, where the nova eruption causes the binary to slightly disconnect and for the hot white dwarf to drive a substantially higher accretion rate (Shara et al. 1986; Shara 1989).  The mechanisms and prevalence of hibernation are under much continuing discussion (Naylor et al. 1992; Evans et al. 2002; Martin \& Tout 2005; Warner 2006; Shara et al. 2007; Thorstensen et al. 2009).  A necessary part of hibernation is that the post-eruption nova will slowly decline in brightness as the white dwarf cools (and the accretion rate falls).  This cooling has estimated rates of perhaps a tenth of a magnitude in the first decade or century after the eruption ends.  These predicted declines have been sought for many novae (e.g., Duerbeck 1992).  The only novae with known systematic and significant declines are V1500 Cyg (Sommers \& Naylor 1999), RW UMi (Bianchini et al. 2003; Tamburini et al. 2007), T Pyx (Schaefer 2005; 2010), and CK Vul (Shara \& Moffat 1982; Shara et al. 1985; Hajduk et al. 2007).  These declines all have long times scales of one to two centuries or more.

In this paper, we will take these various mechanisms (SSS luminosity, irradiation of the companion, plus hibernation and decline) and work them together as cause and effect.  In particular, we will identify a specific group of novae that all share five uncommon properties.  We then identify the physics processes that make this group a distinct nova subset and why the suite of five uncommon properties are causally connected.

\section{Novae with High $\Delta$m}

Recently, we have been measuring the brightness of many novae from {\it before} their eruptions (Collazzi et al. 2009), with this necessarily being made on archival photographic plates.  We measured pre-eruption magnitudes ($m_{pre}$) and post-eruption magnitudes ($m_{post}$) for 31 classical novae and for 19 recurrent nova eruptions.  An important parameter is the difference between the brightness level across the eruption, $\Delta$m=$m_{pre}-m_{post}$, where a positive $\Delta$m indicates that the post-eruption quiescent level is brighter than the pre-eruption quiescent level.  If the nova eruption has no significant effect on the long-term accretion rate, then $\Delta$m$\approx$0 mag.  Surprisingly, a number of novae were found with high-$\Delta$m.

A typical case is illustrated in Figure 1, which shows pre-eruption and post-eruption images from the Palomar Sky Survey of RW UMi.  This nova went off in 1956.  The first epoch Palomar Sky Survey plates were taken four years before and both showed that the nova was below the plate limits with B$>$21.2 (Kukarkin 1963).  However, the second epoch Palomar Sky Survey plates show RW UMi to be moderately bright at B=18.33 (Collazzi et al. 2009).  The second epoch plates were taken 33 years after the eruption, so there is no plausible case that we are seeing the declining tail of the thermonuclear runaway.  Bianchini et al. (2003) and Tamburini et al. (2007) have found that RW UMi is also fading in brightness at a very slow rate.  So RW UMi is an example of a high-$\Delta$m nova with a slow post-eruption decline.

Collazzi et al. (2009) have collected five high-$\Delta$m nova.  In this paper, we add further examples of GQ Mus, CP Pup, and T Pyx.  The eight high-$\Delta$m novae are V723 Cas, V1500 Cyg, V1974 Cyg, GQ Mus, CP Pup, T Pyx, V4633 Sgr, and RW UMi.

We have collected all available light curve data and plotted these in Figures 2-9 for all eight high-$\Delta$m stars.  The data sources are cited in the figure captions.  We have included details on the measured $\Delta$m values in Table 1.  In all eight systems, we have pre-eruption magnitudes, or limits, and these are represented in the figures by a horizontal line continuing to the post-eruption interval.  In all eight systems, the post-eruption light curve has flattened out, indicating the end of the eruption.  In all eight systems, the post-eruption light curve is far brighter than the pre-eruption level, and this is just saying that the $\Delta$m values are large and positive.  These light curves provide an evocative and compelling case for the existence of high-$\Delta$m stars.

For those for which it can be measured, {\it all} of the high-$\Delta$m novae have a very slow decline after the light curve has flattened off.  In principle, we should wonder whether these high-$\Delta$m tails are just the last stages of the ordinary declining tail of an eruption.  (These declines also make the measure of $\Delta$m a time dependent quantity, a problem that is not serious as any introduced uncertainty is only a small fraction of $\Delta$m.)  In practice, the standard definition of the end of the eruption is the time when the decline in the light curve effectively stops.  Strope et al. (2010) find that the late light curves are lines with sharp breaks to a flat quiescence when plotted with logarithmic time, and this readily identifies the date of the end of the eruption.  The post-eruption declines have slopes of 0.016-0.09 mag yr$^{-1}$, and these are comparable with the slopes expected from the hibernation model, so these small declines (visible only over decades) are effectively flat when compared to all slopes during eruption.  Another strong reason to know that the eruptions are finished is that the nova systems show no spectroscopic or photometric evidence of residual hydrogen burning at the late times.  Another way of saying this is to point out that the current light curves represent the nova so long after eruption that no one would care to say that the eruption is continuing so long.  For example, V1500 Cyg is a very well observed system, and it is the {\it fastest} of all classical novae in eruption, and so it is unbelievable that the white dwarf is still in an `eruption' even a decade after its 1975 peak, much less still in eruption in the year 2009.  The thermonuclear runaway has long stopped and its ashes have long cooled.  In all, we accept the traditional definition that the eruption ends when the light curve goes essentially flat.

Between Collazzi et al. (2009) and this paper, we have measures of $\Delta$m for 53 eruptions, with a histogram presented in Figure 10.  We see an approximately Gaussian distribution with a mean at 0.0 mag and standard deviation of 0.4 mag, plus eight outliers with large $\Delta$m.  Novae in quiescence vary substantially in brightness on all time scales (Schaefer 2010; Kafka \& Honeycutt 2004; Honeycutt et al. 1998), so we should expect that $\Delta$m will vary by perhaps half a magnitude even for time-averaged magnitudes.  So the Gaussian distribution centered on $\Delta$m=0 is just the expected case where the nova event does not change the quiescent state.  The eight high-$\Delta$m stars are clear outliers to the distribution of ordinary novae.

These high-$\Delta$m novae are fairly startling, because the standard picture of novae is that $\Delta$m$\approx$0 mag (Robinson 1975; Warner 1989; 1995; 2002; 2008).  The measurement errors are greatly smaller than the deviations from zero for the outliers.  The ordinary variations of novae on all time scales can cause deviations from zero, but these variations in $\Delta$m (with RMS of 0.4 mag as seen for the ordinary novae in Figure 10) are also much smaller than deviations from zero for the high-$\Delta$m cases.  Most of the high-$\Delta$m novae are greater than three-sigma outliers, as the original compilation of Collazzi et al. (2009), selected without any bias for or against novae with large $\Delta$m, has the three-sigma limit at 2.7 mag.  The deviations from zero are all (or almost all) greater than 2.5 mag (a factor of ten in brightness), which corresponds to greater than a factor of ten rise in the accretion rate, so these are indicative of important changes in the systems.  In all, we are left with a result that a moderate fraction of novae have a significant and startling post-eruption brightening.

\section{Other Properties of the High-$\Delta$m Novae}

To search for commonalities amongst the eight high-$\Delta$m novae, we have collected from the literature a variety of properties for all eight novae.  In particular, we are interested in the orbital period ($P_{orb}$), whether the system has a long-lasting SSS, whether the white dwarf (WD) has a high magnetic field (i.e., whether it is a polar or intermediate polar), and whether the post-eruption light curve shows a significant and systematic decline in brightness.  In Table 2, we collect information about the various fundamental properties, the $\Delta$m values from Table 1, as well as on $P_{orb}$, the WD magnetic field, and post-eruption declines.  The third column gives the light curve class in hte new system presented by Strope et al. (2010), where the `S' indicates a smooth decline, `P' indicates a smooth decline with a plateau, `J' indicates a light curve with many jitters or flares, and `U' indicates an unknown light curve class due to a poorly sampled light curve, while the $t_3$ value in days is then given in parentheses.  Full references and explanations are given in the remainder of this section.  We see that the high-$\Delta$m novae generally have short orbital periods, long-lasting SSS, highly magnetized white dwarfs, and long slow post-eruption declines.

All novae should become SSS when their photosphere has receded enough so as to reveal the hydrogen burning near the surface of the white dwarf.  In general, this SSS will then soon turn off when the hydrogen burning stops.  Hachisu \& Kato (2010) present a theoretical relation that the SSS is expected to turn off at a time $T_{th}$=5.3$t_3^{1.5}$ days after eruption (for $8 \lesssim t_3 \lesssim 80$ days where $t_3$ is the time it takes the light curve to decline by three magnitudes from peak).  A fast nova with $t_3=4$ (like V1500 Cyg) should have the SSS turn off perhaps around 40 days after the peak, while a moderate nova with $t_3=45$ days (like GQ Mus) should turn off after 4.4 years.  For a minority of novae, this theoretical prediction can be compared to the directly observed X-ray turn off time ($T_{obs}$).  Values of $t_3$, $T_{th}$, and $T_{obs}$ are listed in Table 2 for all eight novae.  (For the novae with $t_3$ outside the prescribed range, we still quote turn off times as based on the formula, as these values will still be plausible approximations of the expected time scales for $T_{th}$.)  If a nova SSS lasts substantially longer than $T_{th}$ (or at least longer than 5 years), then we label the nova as a `long-lasting SSS'.  This definition requires X-ray observations sufficiently long after the hydrogen burning stops but before the extended SSS stops, and few novae have the necessary data to test for this property.  Alternatively, the existence of the long-lasting SSS might be demonstrated without the use of X-ray data, perhaps by the observation of irradiation on the companion star.  We are disturbed that the quintessential nova SSS V1974 Cyg is not included by our definition, because its X-ray turn-off is well observed to be around 700 days (Krautter et al. 1996), which is fully consistent with ordinary hydrogen burning with no additional energy source (Hachisu \& Kato 2006).  However, this nova is the most luminous known SSS (Ness et al. 2007), so we wonder whether the luminosity can be somehow traded off with the duration so as to help produce the effects associated with a high $\Delta$m.  In particular, maybe we should be using the total fluence from the SSS (a much harder quantity to measure) such that with some latency, the companion star have a higher accretion rate driven for a substantial time after the SSS turns off.

\subsection{V723 Cas}

V723 Cas brightened up to V=7.1 in 1995.  Formally, the nova declines by two magnitudes from its peak in 263 days, and fades by three magnitudes from peak with $t_3$=299 days (Strope et al. 2010), but this is ill-defined because the light curve shows prominent jitters up and down around the peak.  Collazzi et al. (2010) reports on a pre-eruption magnitude from the USNO B2 magnitude from 1985 to be B=18.76.  The magnitude in both 2003.80 and 2009.95 is V=15.25, so apparently the light curve has gone flat and the eruption ended around late 2003.  The light curve was also see to go flat at B=15.75 by Goranskij et al. (2007).

The eruption ended in late 2003, but V723 Cas still shows X-ray activity that was not there before the eruption (Ness et al. 2008).  This nova has remained a ``bright super soft source 12 years after outburst" (Ness et al. 2008).  Observations with the {\it Swift} XRT detectors shows V723 Cas to remain bright as of June 2009.  This long-lasting SSS is understandable within the theoretical model of Hachisu \& Kato (2010) as being due to burning of  residual hydrogen that takes a long time to die out on a slow nova.  Alternatively, Ness et al. (2008) advance the idea that the burning hydrogen is not residual from the eruption, but rather supplied by the continuing accretion, such that the long-lasting SSS will keep going long after any residual hydrogen is burnt up.  Indeed, as with all the other high-$\Delta$m stars, the post-eruption brightness level is elevated in the optical and that can only mean higher accretion rates that must somehow be associated with the eruption.  With this activity after the end of the eruption, alternative definitions might be considered for determining the end of the eruption.  While it is possible to think up non-standard definitions, it really does not matter what we call it, as this paper is simply pointing to a time interval after the light curve has gone essentially flat during which a fraction of novae display a much brighter light than before the eruption, and during which this same fraction also shows a characteristic suite of uncommon properties.  By whatever the name or cause for the time interval, these high-$\Delta$m novae form a distinct subset that we are trying to understand in this paper. 

V723 Cas has an orbital period of 16.62 hours (Goranskij et al. 2007).  However, the nova also shows a significant periodicity at 15.24 hours (Cochol \& Pribulla 1998).  The existence of multiple periodicities is a hallmark of intermediate polar systems with a high magnetic field on the WD.

\subsection{V1500 Cyg}

V1500 Cyg is one of the brighter novae seen, with a very well observed peak at V=1.9 mag in 1975.  The pre-eruption light curve (Collazzi et al. 2009) has a number of deep limits and two positive detections on old Palomar plates, with B=21.5 and V=20.5.  The decline was very fast, with $t_3$=4 days (Strope et al. 2010), making it the all-time fastest known classical nova eruption.  When the light curve (c.f. Figure 3) is plotted versus logarithmic time since the eruption, the linear late decline has a sharp break to a nearly flat slope around the year 1982-1984 (7-9 years after the eruption), and so this is the time when the eruption ended.  Ever since this end, the brightness level has been greatly brighter than the pre-eruption level, so V1500 Cyg has a high-$\Delta$m.  From $\sim$1982 to 2009, V1500 Cyg has been undergoing a steady and slow decline at a rate of 0.09 mag yr$^{-1}$.  At this rate, V1500 Cyg will return to its pre-eruption level around the year 2035, sixty years after its eruption.

The orbital period of V1500 Cyg is 3.35 hours (Semeniuk et al. 1995).  However, the polarimetric period (tied to the spin of the WD) is 1.8\% shorter than the photometric period (tied to the orbit), so this is a sure sign that the system is a polar system, with the slight inequality presumably being caused by the eruption (Stockman et al. 1988, Schmidt \& Stockman 1991).  V1500 Cyg was the first discovered nova that was a polar.

V1500 Cyg has a galactic latitude of $-0.07\degr$, and therefore the extinction is so high that any supersoft X-ray source could not be detected.  Nevertheless, optical observations of the companion star provide convincing evidence that the WD is a highly luminous source of high energy photons.  In particular, Schmidt et al. (1995) find that the inner hemisphere of the companion star has a surface temperature of 8000 K while the outer hemisphere has a surface temperature of 3000 K, with the difference caused by the WD emitting high energy photons with a luminosity of $\sim$5 $L_{\sun}$ in high energy photons.  They made a plausible assumption that this luminosity was blackbody light coming uniformly from the entire surface of the WD, for which they then calculated a temperature of $\sim$100,000 K.  With this assumption, the WD luminosity is too cold to be called a SSS.  However, a more likely possibility is that the luminosity comes from a small polar cap where the accretion and the fresh hydrogen are arriving.  After all, V1500 Cyg is the original polar nova, for which the fraction of the WD surface covered by the pole caps is $f$$\sim$0.01.  If the luminosity comes from a small area, then any blackbody temperature ($T$) must be substantially hotter than 100,000 K so as to provide the $\sim$5 $L_{\sun}$.  This known luminosity will scale as $fT^4$, so that for $f\sim$0.01, we get $T\sim300,000$ K, and this temperature is hot enough to be considered a SSS.  So we know confidently that the nova eruption triggered the WD to be a highly luminous emitter of high energy photons that is long-lasting, and the temperature of this radiation is consistent with that of a long-lasting SSS.

\subsection{V1974 Cyg}

V1974 Cyg reached a peak of V=4.3 in early 1992.  The only useful pre-eruption data are the {\it limits} from the two epochs of the Palomar sky survey, with B$>$21 (Collazzi et al. 2009).  The nova declined with $t_3$=43 days, and there is a plateau in the light curve starting 9 mags below the peak (Strope et al. 2010).  The light curve (see Figure 4) has flattened off by 2002.  After 2002, the light curve has no apparent decline, but there are too few points and too short a time interval to put fine constraints on this.  The post-eruption color is $B-V\approx$0.25 mag (Shugarov et al. 2002) so the late-time magnitude is $V=16.63$ and $B=16.88$, for $\Delta$m$>$4.12 mag.

V1974 Cyg has an orbital period of 1.95 hours (Shugarov et al. 2002), which is {\it below} the period gap.  The system has two photometric periods, with this being the hallmark of highly magnetized WDs (De Young \& Schmidt 1994; Semeniuk et al. 1995a).  Chochol et al. (1997) are pointing to features in the nova shell that they interpret as `magnetic fountains', with the implication that the WD is highly magnetized.

V1974 Cyg was the second discovered long-lasting SSS (Krautter et al. 1996), and it remains as the most luminous known of all SSS novae (Ness et al. 2007).  It lasted until an X-ray turn-off time of 600 days after the eruption, and it was last seen about 700 days after the eruption.  With its observed $t_3$=43 days, the Hachisu \& Kato (2010) model suggests that ordinary burning of residual hydrogen should continue with a turn-off time of around 4.1 years after the eruption.  Within this model, the duration of the long-lasting SSS of V1974 Cyg is `normal', however the extreme luminosity demonstrates that something unusual is going on with the long-lasting SSS.  While V1974 Cyg does not fit our formal definition for a `long-lasting SSS', we think it significant for the physics of the situation that it is by far the most luminous known SSS nova.  Perhaps the relevant quantity is not the total duration of the SSS phase (what we are considering in this paper), but rather the total irradiation onto the companion star (a hard quantity to measure), such that V1974 Cyg would fit in well with the other V1500 Cyg stars.

\subsection{GQ Mus}

GQ Mus came to a peak of V=7.2 mag in early 1983, followed by a decline with $t_3$=45 days (Shafter 1997) or $t_3$=48 days (Whitelock et al. 1984).  The pre-eruption level was measured to be $V=21$ as based on the SRC J sky survey plates (Krautter et al. 1994).  The light curve (Figure 5) goes flat in 1988.  In 1994, Diaz et al. (1994) give average magnitudes of B=17.62 and R=17.50 (as calibrated with the information in Diaz \& Steiner 1989), for which we take V=17.55.  This gives $\Delta$m=3.45 mag.   The nova shows a slow steady decline from 1988 to 2008 at the rate of 0.08 mag yr${-1}$.  At this rate, GQ Mus will return to is pre-eruption level around the year 2050, 67 years after its eruption.

GQ Mus was the original long-lasting SSS (\"{O}gelman et al 1993; Diaz et al. 1995).  It was detected with {\it ROSAT} in 1992, and had finally turned off between January and September 1993, for a total duration of a bit over 10 years (Shanley et al. 1995).  If we apply the Hachisu \& Kato (2010) model with $t_3$=45 days, then the predicted X-ray turn off is 4.4 years after the eruption.  Thus, GQ Mus is a long-lasting SSS that lasted around six years longer than predicted.

The orbital period of GQ Mus is 1.42 hours (Diaz \& Steiner 1994), with this being the smallest known period for novae.  Diaz \& Steiner (1994) used a detailed analysis of red and blue light curves and emission lines to demonstrate that GQ Mus is a polar.

\subsection{CP Pup}

CP Pup was one of the all time brightest novae, reaching V=0.7 in late 1942.  The decline was very fast, with $t_3$=8 days (Strope et al. 2010).  From our constructed light curve (Figure 6), the light curve goes nearly flat sometime from 1960 to 1970.  From 1971 to 2009, the light curve has a slow steady fade from 14.24 to 15.19, a rate of 0.026 mag yr$^{-1}$.

Gaposhkin (1946) tells briefly that he searched for the pre-eruption nova on the Harvard plates and found none to B=17.  While this is undoubtedly approximately correct, we know that all old comparison sequences have problems for faint stars that are typically 1.0 magnitude (Schaefer 1994, 1995, 1996, 1998; Schaefer et al. 2008).  To get a real limit on a modern magnitude scale, we have both calibrated a sequence of comparison stars and have made a new search of the deep plates at Harvard.  Our comparison stars were calibrated in both B-band and V-band with the SMARTS 1.3-m telescope on Cerro Tololo on two photometric nights.  Our comparison sequence has 12 stars from 14.09 to 20.45 in the B-band.  On these two nights, we also got measures of CP Pup itself, with B=15.38 and V=15.32 on JD2455094.88, while B=15.45 and V=15.19 on JD2455120.84.  We have travelled to Harvard to examine all the original deep plates from before the eruption.  CP Pup was not seen on any plates.  Our deepest limit (as calibrated from the faintest stars in our comparison sequence that were visible) is B=19.4.  Thus, we now have a well-measured limit on the pre-eruption brightness of B$>$19.4.  For a color in quiescence of $B-V=0.2$ mag (Diaz \& Steiner 1991, and from our measures in 2009), and with the observed V=14.24 in 1971, we get B=14.44 as the post-eruption magnitude.  Then, $\Delta$m$>$4.96 mag.

With the eruption of CP Pup being so long before the invention of the X-ray telescope, we have no way of directly measuring the presence of a long-lasting SSS after the eruption.  The earliest X-ray observation of CP Pup was in 1993 (Balman et al. 1995), 51 years after the eruption.  Their {\it ROSAT} observation detected the nova from 0.1-2.4 keV, but the inferred temperature (kT$\ge$1 keV) is not consistent with a SSS.  Orio et al. (2009) found similar results with {\it ASCA} in 1998 and with {\it XMM-Newton} in 2005.  Observations at such late dates have little real chance of detecting any long-lasting SSS, because the supersoft sources are seen out only to 14 years after eruption (for V723 Cas) and have decline timescales of 60 years (for V1500 Cyg).  So we conclude that we have no useful information as to whether CP Pup had a long-lasting SSS or not.

CP Pup has an orbital period of 1.47 hours (O'Donoghue et al. 1989), which is one of the fastest period known for a nova.  CP Pup is an intermediate polar system (Balman et al. 1995; Orio et al. 2009; Warner 1985) with a highly magnetized WD.

\subsection{T Pyx}

T Pyx is one of only ten known recurrent novae in our Milky Way galaxy, with observed eruptions in 1890, 1902, 1920, 1944, and 1967 (Schaefer 2010).  Schaefer (2010) has exhaustively measured and collected all the world's data on T Pyx in eruption, and finds that all the light curves are consistent with a single template.  The peak magnitude is V=6.4 mag, while the decline has $t_3$=62 days.

T Pyx is fading slowly over the last 120 years (Schaefer 2005; 2010).  The quiescent B-band brightness has gone from 13.8 mag {\it before} the 1890 eruption to 15.7 mag in 2009 (Figure 7).  The five nova events are brief and have no apparent effect on this secular decline.  This decline has an average rate of  0.016 mag yr$^{-1}$ and corresponds to a decrease by a factor of 30 in the quiescence accretion rate in the system.

The orbital period of T Pyx was discovered to be 1.83 hours (Schaefer et al. 1992).  This period was confirmed by Patterson et al. (1998), who also found a second significant photometric periodicity (2.63 hours) whose amplitude was time variable.  The stability of the primary photometric periodicity over several decades guarantees that this is associated with the orbital period.  Recently, Uthas (2009) has spectroscopically confirmed the orbital period of 1.83 hours.

The presence of {\it two} nearly equal photometric periods is the hallmark of magnetic WD systems.  The significant offset between the orbital period (1.83 hours) and the WD rotation period (2.63 hours) is characteristic of intermediate polars.  T Pyx is certainly not a polar because its spectral energy distribution shows a $\nu^{1/3}$ power law characteristic of an accretion disk (Schaefer 2010).  Despite not having, say, polarimetric data, as with the other stars, we will take the existence of two photometric periodicities as reasonable evidence that the WD is highly magnetic. 

A key discovery for understanding T Pyx is that it had an ordinary nova eruption in the year 1866$\pm$5 (Schaefer et al. 2010).  The basis of this a series of {\it HST} images that show the nova shell around T Pyx to be homologously expanding with a velocity of 500-715 km s$^{-1}$.  The fact that all the knots are expanding with the same fractional rate is proof that the motion has suffered no significant deceleration (because knots of the observed widely varying masses must inevitably be slowed by greatly different amounts and this would lead to greatly different fractional rates, which is not observed).  With no significant deceleration, the (distance independent) eruption date can be determined simply by backtracking the motion of the knots, with this date being 1866$\pm$5.  With no significant deceleration, we realize that the 1866 eruption could not have been a recurrent nova event, both because recurrent novae all have expansion velocities of $FWHM\ge$2000 km s$^{-1}$ (Pagnotta et al. 2009) and because the observed expansion velocity in the 1967 eruption was 2000 km s$^{-1}$ (Catchpole 1969).  The low expansion velocity of the 1866 shell ($\approx600$ km s$^{-1}$) is characteristic of an ordinary nova eruption.  Further, Schaefer et al. (2010) used the $H\alpha$ flux to estimate the mass of the 1866 ejecta as $\sim$10$^{-4.5}$ M$_{\odot}$, and this is certainly several orders of magnitude larger than is possible from a recurrent nova event while being typical of ordinary nova events.  Thus, the 1866 eruption was an ordinary nova eruption, quite different from the later recurrent nova events.  To get the observed mass of ejected material from a near-Chandrasekhar mass WD, the accretion rate must be low at around $4\times10^{-11}$ M$_{\odot}$ yr$^{-1}$ (Nomoto 1982; Townsley \& Bildsten 2005; Shen \& Bildsten 2007), exactly as expected for the system when the accretion is being driven only by angular momentum losses associated with gravitational radiation (Patterson 1984).  For this accretion rate, Schaefer et al. (2010) calculate that the T Pyx system would have been B=18.5 mag.  Thus, the $\Delta$m value across the 1866 ordinary nova eruption would have been 18.5-13.8=4.7 mag.  In this paper, we are pointing to the 1866$\pm$5 eruption of T Pyx as the high-$\Delta$m event.  

The 1866 ordinary nova eruption occurred so long ago that there is no way to directly test whether a long-lasting SSS existed.  T Pyx was only a weak and hard X-ray source in the year 2006 as viewed with {\it XMM-Newton} (Selvelli et al. 2008), but this says little about any long-lasting SSS from either the 1967 or 1866 eruptions.  Indirect evidence for a long-lasting hot and luminous WD comes from the shape of the orbital modulation, the blue color, the high-excitation spectrum, as well as the high visual and bolometric luminosity (Patterson et al. 1998).  A very hot and luminous white dwarf provides the only known solution to a deep and fundamental problem for T Pyx.  This problem is that T Pyx currently has an accretion rate of $>10^{-8}$ M$_{\odot}$ yr$^{-1}$ (Schaefer et al. 2010; Selvelli et al. 2008; Patterson et al. 1998) and $\sim30\times$ larger back in 1890, and the very high accretion rate is `impossible' for any cataclysmic variable star with a period below the period gap for which the only apparent mechanism for angular momentum loss is gravitational radiation.  Knigge et al. (2000) propose a solution to this paradox is that the white dwarf in T Pyx is so hot and luminous that its irradiation of the nearby companion star drives the high observed accretion rate.  They propose a mechanism where the irradiation drives a stellar wind on the companion star, while Schaefer et al. (2010) propose a mechanism where the irradiation puffs up the stellar atmosphere of the companion which drives the high accretion.  Detailed calculations in the two papers show that both mechanisms should be operating, so the only question is which one dominates.  With no other solution to the basic paradox (how T Pyx can have such a high accretion rate) and the reasonableness of the Knigge et al. solution, we conclude that T Pyx has a very hot and luminous WD.  Given the secular decline in brightness since 1890, the WD must have been much hotter and more luminous a century ago.

\subsection{V4633 Sgr}

V4633 Sgr came to a peak of V=7.4 in early 1998.  The pre-eruption nova is known only from {\it limits} from the Palomar sky survey, with B$>$21 (Collazzi et al. 2009).  The decline has $t_3$=44 days (Strope et al. 2010).  The light curves declines until 2006, after which it remains constant at B=18.7 and V=18.7.  This makes the $\Delta$m value $>2.3$ mag.

V4633 Sgr was observed 934, 1083, and 1265 days after eruption from 0.2-10 keV with the {\it XMM-Newton} satellite (Hernanz \& Sala 2007).  The nova was easily detected on all three dates, with little variation between the dates.  A hard component was certainly visible, with the likely origin being shock heated ejecta in the expanding shell.  Hernanz \& Sala (2007) state that residual hydrogen burning (which would create a SSS) had already turned off within 2.6 years after optical maximum, but only because their model would then suggest a large photospheric size.  On the other hand, their best fit three-temperature spectral models have a high luminosity from a component with $kT\approx0.07$ keV (800,000 K) in all epochs of observation.  With this, V4633 Sgr is a SSS lasting for at least 3.6 years with no sign of turn off.  This could be compared to the Hachisu \& Kato prediction from the $t_3$ value that the X-ray turn off will be 4.2 years after the eruption.  With the uncertainties in the observations as well as in what is accepted as a long-lasting SSS, we conclude that it is ambiguous whether V4633 Sgr had one.

The orbital period of V4633 Sgr is 3.01 hours (Lipkin et al. 2001), which is just at the top edge of the period gap.  Lipkin \& Leibowitz (2008) discovered a second photometric period at 3.09 hours that was time variable and they made the usual conclusion that V4633 Sgr is a slightly asynchronous polar system, where the 3.09 hour WD rotation period was knocked out of synchronism by the eruption.

\subsection{RW UMi}

RW UMi was only discovered in 1963 (Kukarkin 1963) long after its peak in 1956.  Archival plates in Russia and at Soneberg Observatory sketched out a poorly sampled light curve (Figure 9) with a peak at least as bright as B=6.  Kukarkin (1963) and Collazzi et al. (2009) use the Palomar sky survey plates (Figure 1) to show that the pre-nova was fainter than B=21 mag.  The decline is estimated to have $t_3$=140 days (Shafter 1997).  Apparently, the light curve went to quiescence sometime around 1965.  Szkody (1994) gives B=18.46 and V=18.52 in 1989.  With this, we get $\Delta$m$>$2.52 mag.  Bianchini et al. (2003) has collected all the magnitudes from 1986-2002 and concluded that RW UMi is fading steadily at the slow rate of $\sim0.02$ mag yr$^{-1}$.  Tamburini et al. (2007) have updated this to 2006, and they conclude that the decline rate is $\sim0.03$ mag yr$^{-1}$. 

The orbital period of RW UMi is 1.42 hours (Retter \& Lipkin 2001), in a tie with GQ Mus as the fastest known orbital period for any nova.  Tamburini et al. (2007) find two photometric periodicities and make the usual conclusion that RW UMi is an intermediate polar system.

RW UMi has never been looked at in the X-ray regime.  With an eruption in 1956, any modern X-ray observation would have little bearing on whether RW UMi has a long-lasting SSS.

\section{Connections Between Properties}

We have just identified a class of novae defined by a large $\Delta$m.  While investigating these eight novae, we were struck by the similarity of their properties.  In particular, most of these eight nova have a coherent set of properties; including (a) a significantly high $\Delta$m, (b) short orbital periods near or below the period gap, (c) continuing supersoft emission long after the eruption is over, (d) magnetic WDs, and (e) slowly declining optical brightness after the end of the eruption.  Each of these properties is uncommon.

Few novae have a significantly different pre-eruption and post-eruption brightnesses.  Here, with novae jittering and flickering around on all time scales (Collazzi et al. 2009), we adopt a threshold of $\Delta m = 1.5$ mag as being a significant difference.  In a study of $\Delta$m for 37 novae, Collazzi et al. (2009) found that only 5 (14\%) had $\Delta m > 1.5$ mag.  The selection of novae for this study was not biased towards or against high values for $\Delta m$.  In our work here, we have added three more high-$\Delta m$ cases.

Most novae have orbital periods longer than the period gap ($2<P_{orb}<3$ hours).  Diaz \& Bruch (1999) provide a study of the orbital period distribution for classical novae.  At the time, of 31 novae with known periods, only 4 novae have $P_{orb}<2$ hours.  The fraction of novae below the period gap is thus 13\%.  The four novae below the gap are GQ Mus, CP Pup, RW UMi, and V1974 Cyg, all of which are in our new high-$\Delta m$ group.  Since that time, the orbital period of T Pyx has also been found to be below the period gap, and T Pyx is also a member of our new group.

The statistics of novae with long-lasting supersoft emission is somewhat ill-defined, as the X-ray observations are usually poorly-sampled in time.  Orio et al. (2001) used ROSAT to look at 30 novae within ten years of the eruptions and found only 3 SSS (V1974 Cyg, GQ Mus, and Nova LMC 1995), for 8\%.  The fraction of novae with long-lasting SSS would likely be larger than this if the X-ray observations had been well-sampled for all 30 novae.  Ness et al. (2007) report on a survey of classical novae with the {\it Swift} X-ray detectors, for which only three novae have been looked at more than 100 days after peak (to give the SSS time to turn on), from their recent sample (to avoid old novae selected for their X-ray behavior), and with column density less than $10^{22}$ cm$^{-2}$ (so that a SSS could be seen), with one of the three (V574 Pup) having a SSS.  With poor statistics, this gives a fraction of one-third of the novae develop long-lasting SSS, or at least that happen to have been observed.  With an uncertainty of perhaps a factor of two, we take the fraction of nova having long-lasting supersoft emission to be $\sim25\%$.

The fraction of high magnetic field white dwarfs in novae is also poorly known.  In their invited review on magnetic white dwarfs, Wickramasinghe \& Ferrario (2000) found that $\sim5\%$ of isolated white dwarfs and $\sim25\%$ of white dwarfs in cataclysmic variables are magnetic.  Their lists include only one magnetic nova (V1500 Cyg), but many more have since been discovered.  Townsley \& Bildsten (2005) use a population model to suggest that the magnetic fraction for novae is comparable to that of other cataclysmic variables.  In a survey of the literature, we found 22 classical novae which have been identified as being polars or intermediate-polars, while perhaps 100 classical novae have been observed closely enough such that magnetic behavior would have been recognized.  We will take the fraction of novae with high magnetic field white dwarfs to be a conservative 25\%.

Novae in quiescence vary in brightness on all time scales, and this makes it difficult to identify those with long-running, significant, and systematic post-eruption declines.  Many novae have adequate long-term light curves (Collazzi et al. 2009; Schaefer 2005; 2010; Honeycutt et al. 1998; Kafka \& Honeycutt 2004; Duerbeck 1992; plus many other publications for individual novae), and we have examined roughly 30 novae quiescent light curves with excellent coverage over many decades in the data base of the American Association of Variable Star Observers (AAVSO).  In all, we estimate that roughly a hundred old novae have been examined closely enough in quiescence such that a steady decline in brightness would have been detected.  Out of these, we know of only five observed cases (V1500 Cyg, GQ Mus, CP Pup, T Pyx, and RW UMi) with systematic  and significant declines.  Thus, the fraction of novae with declines is of order 5\%. 

What is the probability that any two uncommon properties would predominate in one group of stars without a causal connection?  Our high-$\Delta m$ group has been chosen for one of these uncommon properties, so what is the probability that some number of these stars will share an uncommon property with a frequency $f$?  The probability is $f^{8}$ that all 8 of the stars have the uncommon property, and is $8(1-f)f^{7}$ that exactly 7 of the 8 stars have the uncommon property, and so on as given by the binomial theorem.  Such probabilities have to be multiplied by the number of trials, here represented by the number of properties ($P_{orb}$, white dwarf mass, nova shells, ...) and the number of distinct cases ($P_{orb}<2$ hours, $P_{orb}>1$ day, ...) that could have produced a correlation.  This number of trials is difficult to determine, but appears to be between 10 and 100.  If after correction for the number of trials, the resultant probability is sufficiently low, then we can reject the idea that the clustering is by randomness, so we can conclude that there must be some sort of causal connection.

The high-$\Delta m$ novae have 5-out-of-8 with orbital periods below the period gap ($P_{orb} \leq 2$ hours), a property that happens in 13\% of all novae.  This is improbable at the 0.0015 level, which is beyond the usual three-sigma confidence level.  Nevertheless, after multiplying by 10-100 for the number of trials, we are left without a significant connection.  Alternatively, we can ask what is the probability that 5-out-of-5 novae with known periods below the gap are in our new high-$\Delta m$ group  (which constitutes 14\% of novae)?  This probability is $0.14^5=0.000054$.  For any plausible number of trials, the final probability is better than the three-sigma confidence threshold.  Thus, the connection between high-$\Delta m$ and short-$P_{orb}$ looks likely.

The high-$\Delta m$ novae have 4-out-of-5 (which have a confident determination) with long-lasting supersoft emission, a property that happens in $\sim25\%$ of novae.  The probability of this is 0.016, which is not even at the three-sigma confidence level even before being multiplied by the number of trials.

The high-$\Delta m$ novae have 8-out-of-8 cases with high magnetic fields on their white dwarfs, with approximately 25\% of all novae being in these categories.  The probability is $0.25^8=0.000015$, which is significant even after multiplying by the number of trials.  So there is some causal connection between high-$\Delta m$ and the white dwarf magnetic field.

The high-$\Delta m$ novae have 5-out-of-5 of those with more than a decade long post-eruption light curves displaying confident systematic declines, with approximately 5\% of novae showing such declines.  The probability is $0.05^5=0.00000031$ that all five novae will show declines if there is no causal connection.  For any plausible number of trials, the final probability is highly significant.  This demonstrates that there must be a causal connection between high-$\Delta m$ and post-eruption declines.

So far, we have been only comparing pairs of properties; the probability of getting a suite of five properties (each uncommon) all coming together in one small group of stars is very low.  It is the {\it joint} occurrence of these properties that makes the group look like a group.  The probability that our group of 8 high-$\Delta m$ stars will preferentially have short orbital periods, long-lasting supersoft emission, high magnetic fields, and declining post-eruption light curves is $0.0015\times0.016\times0.000015
\times0.00000031\approx10^{-16}$.  To get the final probability, we have to multiply by all the number of plausible combinations of properties that we could imagine grouping together. This number of trials is of order 10-100 raised to the fourth power (i.e., $10^4$ to $10^{8}$).  Thus, the final probability, from $10^{-8}$ to $10^{-12}$, is highly significant.  That is, we cannot get such a clustering of uncommon properties by non-causal means.  This constitutes a proof that our suite of five properties have a physical connection for the novae in our new group.

Such statistical proofs as are displayed in this section might be formally satisfying that a group of properties is causally connected.  Nevertheless, some physical connection and explanation is required before this new group will be accepted.  Indeed, without a physical mechanism, the existence of the new group would be of little utility.  So in the next section, we will provide the physical connection and explain why each of the properties is a necessary consequence of the others.

\section{Physical Mechanisms}

What physical mechanisms connect the high-$\Delta m$ events with short orbital periods, long-lasting supersoft emissions, magnetic WDs, and slowly declining post-eruption light curves?  Whenever a nova eruption occurs, the nova light and the ordinary supersoft phase at its end will heat up the companion star.  If the orbital period is short (with the companion star close to the white dwarf), then the surface temperature of the companion star at the inner Lagrangian point will rise substantially, thus pushing a high accretion rate.  If the white dwarf is highly magnetic, then the accretion is funneled onto a small surface area creating a locally very high accretion rate.  With a locally very high accretion rate, the hydrogen falling onto this area will be falling fast enough to allow for a self-sustaining thermonuclear reaction, which will appear as a supersoft source.  This supersoft source will be able to heat the surface of the companion star (especially if the orbital period is short) such that the accretion rate remains high and the supersoft source is largely self-sustaining and will continue for a long time.  With the continuing high accretion rate, the brightness of the accretion disk and hence of the whole system will be much brighter after the eruption than before (hence a high $\Delta m$).  The supersoft sources will not be perfectly self-sustaining, so the luminosity will decline on a time scale of decades to a century, and the accretion and system brightness will decline on a time scale of decades to a century.  Thus, all of our five properties are easily and necessarily connected by well-known physical mechanisms.

Let us now sketch the physics of these connections, partly to demonstrate that the mechanisms work and partly to show that the various properties are indeed connected:

What are the conditions for the sustained nuclear burning of hydrogen on the surface of the white dwarf?  In particular, what accretion rate ($\dot{M}$) is required to produce a sustained SSS?  For accretion onto the entire surface of the white dwarf, the accretion rate required to sustain stable hydrogen burning varies from $3\times10^{-8}$, $1\times10^{-7}$, and $2\times10^{-7}$ M$_{\odot}$ yr$^{-1}$ for white dwarf masses of 0.6, 1.0, and 1.35 M$_{\odot}$ respectively (Shen \& Bildsten 2007; see also Nomoto et al. 2007; Townsley \& Bildsten 2005; Nomoto 1982).  No cataclysmic variable can reach such high rates in a steady state driven by angular momentum loss from even the combination of magnetic breaking and gravitational radiation (Patterson 1984).  However, if a lower accretion rate is funneled onto a small portion of the white dwarf surface (by the white dwarf magnetic field in a polar or intermediate-polar system), then the local accretion rate will be large enough to sustain local stable hydrogen burning.  If for example, the magnetic field funnels the accretion onto only one percent of the white dwarf's surface, then the overall accretion rate will have a stable burning threshold that is 100 times smaller than the rates just quoted.  The fraction of the white dwarf surface covered by the accretion spot ($f$) is measured in various polar systems to be 0.003-0.024 (O'Donoghue et al. 2006), $<$0.015, $\sim0.008$, 0.004 (Schmidt \& Stockman 2001), and 0.018 (Cropper \& Horne 1994); with $f\approx0.01$ representing a typical value.  In general, steady hydrogen burning (and the consequent SSS) requires that the accretion rate be greater than
\begin{equation}
\dot{M}_{SSS} \gtrsim f10^{-7}~{\rm M}_{\odot}~ {\rm yr}^{-1}.
\end{equation}
This means that polar or intermediate-polar systems can achieve a long-lasting supersoft source with a greatly lower accretion rate than novae with non-magnetic white dwarfs.  The accretion rate immediately after the eruption should be largely independent of the magnetic field on the white dwarf, and for novae in general there will be some distribution of $\dot{M}$ which will presumably have high-field white dwarfs being relatively rare.  In such a case, most of the long-lasting supersoft sources will arise on novae with highly magnetized white dwarfs.  This is the physical mechanism that connects the long-lasting supersoft emission with the polars and intermediate-polars.

Can the supersoft source induce a substantially raised temperature on the inner hemisphere of the companion star?  Hydrogen burning will convert 0.7\% of its mass into energy, so the luminosity of the SSS will be
\begin{equation}
L_{SSS}= 0.007 \dot{M} c^2,
\end{equation}
where $c$ is the speed of light.  The luminosity might be somewhat lower than the value from the equation due to incomplete burning of the incoming hydrogen.  For a typical nova with an accretion rate of $10^{-10}$ M$_{\odot}$ yr$^{-1}$ (Patterson 1984), the supersoft luminosity will be $4\times10^{34}$ erg s$^{-1}$.  For the case of polars, this luminosity will shine directly on the inner Lagrangian point, whereas this Roche lobe overflow point will be shadowed by the accretion disk for intermediate-polars and non-magnetic white dwarfs.  We can calculate the surface temperature of the star at the inner Lagrangian point (due to the SSS) with a steady state energy balance, 
\begin{equation}
\sigma T_{SSS}^4 = \sigma T_{unheated}^4 + L_{SSS} / 4 \pi D_{L1}^2.  
\end{equation}
The $D_{L1}$ value is the distance from the white dwarf to the inner Lagrangian point, with $D_{L1} \propto P_{orb}^{2/3}$.  With the $T_{unheated}$ term usually being negligible, we see that $T_{SSS} \propto P_{orb}^{-1/3}$.  This equation does not account for heat flow around the limb of the star, heat flow into the interior of the star, nor shadowing by any accretion disk.  Away from the inner Lagrangian point, the distance from the source will be larger than $D_{L1}$ and there will be an additional factor for the slant angle of the impinging radiation, with the result that the irradiated surface temperature will be substantially smaller than $T_{SSS}$.  For an example with $L_{SSS}=4\times10^{34}$ erg s$^{-1}$ and $T_{unheated}=3000$ K, the $T_{SSS}$ value will vary from 12,000 to 25,000 K as the orbital period decreases from 15 to 1.5 hours.  We can compare this to the specific case of V1500 Cyg, with $P_{orb}=3.35$ hours, $M_{WD}\approx0.9$ M$_{\odot}$, $M_{comp}\approx0.3$ M$_{\odot}$, $L_{SSS}=5$ L$_{\odot}$, and $T_{unheated}=3000$ K (Schmidt et al. 1995).  The predicted $T_{SSS}$ is 17,000 K, which is substantially higher than the observed 8,000 K (Schmidt et al. 1995).  Likely, this difference is easily resolved by ordinary inefficiencies not accounted for in Eq. 3 (like for energy transported away from the local photosphere) as well as the realization that the observed value is averaged over the entire hemisphere while the temperature we calculate (for the inner Lagrangian point alone) will be much hotter.  Nevertheless, the point is that typical conditions will have a SSS greatly heat up the inner hemisphere of the companion star.  Further, we see the connection between the heating of the companion star and the orbital period, with the flux per square centimeter on the companion star rising substantially as the orbital period decreases.  The temperature rise on the companion star is only a moderate power of the orbital period, and the SSS emission is also changed by the accretion rate and the star masses.  (The importance of factors other than period alone is illustrated by the long orbital period of V723 Cas.)  However, Greiner et al. (2003) have already demonstrated that the SSS is strongly correlated with the orbital period.  Thus, we see the physical mechanism which connects the orbital period with the SSS emission.

The irradiation of the companion star by the SSS will substantially increase the accretion rate onto the white dwarf.  Two mechanisms will be operating.  The first mechanism is that the heated companion star will drive a wind, with much of the ejected material falling onto the white dwarf.  This has been addressed in detail by van Teesling \& King (1998) and Knigge et al. (2000).   They calculate that the extra accretion rate will be
\begin{equation}
\dot{M}_{extra}=3\times10^{-7} (R_{comp}/a_{11})[M_{comp}(L_{SSS}/10^{37})]^{0.5} ~{\rm M}_{\odot}~ {\rm yr}^{-1}.
\end{equation} 
Here, $R_{comp}$ and $M_{comp}$ are the radius and mass of the companion star in solar units, and $a_{11}$ is the binary separation ($a$) in units of $10^{11}$ cm.  (For this, we have set $\phi \approx 1$ and $\eta_s \approx 1$ from equation 1 of Knigge et al. 2000.)  To take a typical case, with $M_{WD}=1.0$ M$_{\odot}$, $M_{comp}=0.5$ M$_{\odot}$, and $P_{orb}=2$ hours (so that $a_{11}=0.21$ and $R_{comp}=0.10$), while $\dot{M}=10^{-10}$ M$_{\odot}$ yr$^{-1}$ (so that $L_{SSS}=4\times10^{34}$ erg s$^{-1}$), we have $\dot{M}_{extra}=6 \times 10^{-9}$ M$_{\odot}$ yr$^{-1}$. With this mechanism, the SSS will be able to generate enough accretion to sustain itself.  The second mechanism is that the heated companion star will have its atmosphere expand (puff up) and this will drive a large amount of material over the Roche lobe so as to then fall onto the white dwarf.  The density of the atmosphere will fall off with altitude with an exponential scale height of 
\begin{equation}
H=kT/g\mu m_H, 
\end{equation} 
where $k$ is the Boltzmann constant, $T$ is the surface temperature, $g$ is the surface gravity of the companion star, $\mu$ is the mean atomic weight of the gas, and $m_H$ is the mass of the hydrogen atom.  This scale height will be imposed on the atmosphere down to the depths that the SSS radiation reaches.  The effect of the SSS will be to raise $T$ from $T_{unheated}$ to $T_{SSS}$, greatly increasing the scale height of the stellar atmosphere, and push a high accretion rate. The accretion rate will scale as $\dot{M}_{extra} \propto e^{-\Delta R/H}$, where $\Delta R$ is the radius of the star inside its Roche lobe (Frank et al. 2002, eqn. 4.19).  The application of this equation will require detailed calculations of the radiation deposition as a function of altitude in the atmosphere and knowing the effective altitude of the Roche lobe surface.  Osaki (1985, Eqns. 36 and 30) have performed a similar calculation with the result that 
\begin{equation}
\dot{M}_{extra}=4.4\times10^{-9} \xi(1+M_{comp})^{2/3}~M_{comp}^{2.833} ~{\rm M}_{\odot}~ {\rm yr}^{-1},
\end{equation} 
where
\begin{equation}
\xi=L_{SSS}/(8\pi \sigma a^2 T^4_{unheated}).
\end{equation} 
This result is for the case of a 1 $M_{\odot}$ white dwarf.  For the typical case above, we get $\xi=760$ and $\dot{M}_{extra}=6 \times 10^{-7}$ M$_{\odot}$ yr$^{-1}$.  Despite the uncertainties in these calculations, from either or both of these two mechanisms, the SSS will drive a substantially higher accretion rate after the eruption has ended than before the eruption had started.  With the optical light in nova systems being dominated by the accretion light, we see that the SSS will make $m_{post}$ much brighter than $m_{pre}$.  Thus, we see the physical mechanisms that connect the high-$\Delta m$ events with SSSs.

The SSS drives enough accretion so as to be self-sustaining over long periods of time, but this will not be stable.  In principle, the accretion rate could be fine-tuned so as to achieve exact stability.  However, should the accretion decrease (increase) from this stable case, then the lesser (greater) SSS flux would drive even less (more) accretion, and the system becomes continually fainter (brighter) in a feedback loop.  As all novae have variable accretion, the stable SSS case cannot be maintained.  So a non-self-sustaining SSS is required, with $L_{SSS}$ {\it declining} over time in the relevant cases.  Any theoretical answer to the time scale of the decline will require very detailed calculations of the physics of the accretion onto the white dwarf, the SSS emission, the deposition of the SSS light in the atmosphere of the companion, and the amount of matter pushed over the Roche lobe.  Instead, we can offer an empirical answer.  We see that T Pyx is declining in accretion rate (from 1890 to 2009, $\dot{M}$ has fallen by a factor of over $30\times$), V1500 Cyg is declining in $L_{SSS}$ (with a time scale of 200 years), and RW UMi is declining in brightness (at a rate of 0.3 mag per decade).  So apparently the usual case is that the SSS is not entirely self-sustaining, while both the SSS luminosity and accretion rate are slowly declining on the time scale of a century or so.  Thus, while the physical mechanisms for instability have not been developed in detail, we see that the SSS should generally lead to a declining post-eruption magnitude.

So we now have a set of physical mechanisms that connect all five properties of our groups (high-$\Delta m$, short $P_{orb}$, long-lasting SSS emission, highly magnetized white dwarfs, and declining post-eruption magnitudes).  These mechanisms are well known and expected to be operating.  And, from the previous section, the suite of uncommon properties all residing in a distinct group of stars is very improbable by random chance, so there must be a causal connection.  In all, we can be confident in the existence of our new group of stars as well as the basic causes for why these stars have connected properties.

\section{Discussion}

We have demonstrated that the eight nova form a distinct group with separate properties setting them apart from other novae.  We need a name for this group.  Given our path to recognizing the group, we think of the members as being the `high-$\Delta$m' novae.  But this does not work well as a name because it only highlights one aspect.  After considering descriptive names and acronyms, we can do no better than to follow the tradition of naming the group after the prototype star.  V1500 Cyg makes a perfect prototype, because it confidently has all the key properties, plus it is a well-known and well-studied case.  So we name this subset of novae to be `V1500 Cyg stars'.

Supersoft sources are suspected to be one of the progenitors for Type Ia supernovae (Rappaport et al. 1994; Branch et al. 1995; Hachisu et al. 1999), so we have to ask whether the V1500 Cyg stars will ultimately collapse when the white dwarf is pushed over the Chandrasekhar mass.  For this question, a key set of observations is that at least several of the V1500 Cyg stars are neon novae, including V1500 Cyg (Ferland \& Shields 1978), V1974 Cyg (Hayward et al. 1992), and V723 Cas (Iijima 2006).  These neon nova cannot be progenitors for two reasons.  First, the presence of neon in the ejecta proves that the eruption is dredging up material from the underlying nova (Truran \& Livio 1986), so the mass ejected is more than the mass accreted during each eruption cycle, hence the white dwarf must be losing mass.  Second, the neon indicates that the white dwarf has an oxygen-neon-magnesium composition, and hence does not have the carbon-oxygen composition required to power a supernova. For the progenitor question, another key observation is that the T Pyx brightness and accretion rate are both falling greatly over the last century.  Schaefer et al. (2010) has used this observation to demonstrate that T Pyx is going into hibernation as its accretion rate falls to near-zero, with the subsequent hibernation lasting an estimated 2.6 million years.  The recurrent nova state lasts only for a century or two, so the time scale for any mass increase of the white dwarf is longer than the Hubble time, so such systems cannot produce any useful rate of Type Ia supernovae.  Similar arguments will presumably apply to all the other systems which show long-term declines (V1500 Cyg, GQ Mus, CP Pup, and RW UMi).  Another set of observations is that some of the V1500 Cyg stars do not have enough total mass in the system to get the white dwarf over the Chandrasekhar mass.  Thus, V1500 Cyg has only 0.9+0.2 M$_{\odot}$ (Schmidt et al. 1995), V1974 Cyg has only 0.83+0.2 M$_{\odot}$ (Chochol 1999), and GQ Mus has only 0.7+0.10 M$_{\odot}$ (Hachisu \& Kato 2010).  Finally, there is the theoretical analysis that shows moderate-mass white dwarfs at moderately-low accretion rate will eject more material than they accrete (Yaron et al. 2005), so they cannot accumulate enough mass to become a supernova.  In all, we have multiple strong reasons for knowing that the V1500 Cyg stars are not Type Ia progenitors.

The V1500 Cyg stars show substantial declines in their post-eruption brightness, so we have to understand how this relates to the predictions of the hibernation model.  On the face of it, the V1500 Cyg stars are good demonstrations of the basic mechanism for hibernation (the eruption initiates a high luminosity near the white dwarf that drives a high accretion that then declines over a time scale of a century or so).  So we have found hibernation.  But this cannot be the hibernation scenario as originally envisioned by Shara et al. (1986), because this paradigm requires that all (or most) of the novae take part so as to account for the space density of cataclysmic variables.  The V1500 Cyg stars are only 14\% of the novae, and thus cannot change the space densities appropriately.  The existence of hibernation in the majority of novae remains as an independent issue, for which the discussion (see the many references in Section 1) is still ongoing.  Apparently, the decline to hibernation for most novae is over a time scale of many centuries and is hard to observe because we do not have the required long time spans of measures.  Perhaps we can characterize this majority of nova as having regular hibernation, while the V1500 Cyg stars have `magnetic hibernation' with the SSS making the difference.

~

We thank the observers from the {\it American Association of Variable Stars Observers} for their data as used in our light curves.

~

{}

\begin{deluxetable}{lcccccc}
\tabletypesize{\scriptsize}
\tablewidth{0pc}
\tablecaption{Large $\Delta m$ Novae - Before and After Eruption}
\tablehead{\colhead{Nova} & Band & \colhead{$m_{pre}$} & \colhead{Ref\tablenotemark{a}} & \colhead{$m_{post}$} & \colhead{Ref\tablenotemark{a}} & \colhead{$\Delta m$}}
\startdata
V723 Cas		&	B	&	18.76	&	1	&	15.75	&	2	&	3.01	\\
V1500 Cyg	&	B	&	21.5		&	3	&	17.90	&	4	&	3.6	\\
V1974 Cyg	&	B	& 	$>$21	&	1	&	16.88	&	5, 6	& $>$4.12	\\
GQ Mus		&	V	& $\approx$21	&	7	&	17.55	&	8, 9	& $\approx$3.45	\\
CP Pup		&	B	&	$>$19.4	&	10	&	14.44	&	5, 11	&	$>$4.96	\\
T Pyx 		&	B 	&	18.5		&	12	&	13.8		&	12	&	 4.7  \\
V4633 Sgr	&	B	&	$>$21	&	13	&	18.70	&	14	&	$>$2.3	\\
RW UMi		&	B	&	$>$21	&	13	&	18.46	&	4	&	$>$2.52	\\
\enddata
\tablenotetext{a}{1.	USNO-B1.0 B2 magnitude.  2. Goranskij et al. 2007.  3.	Wade 1987.   4.	Szkody 1994.   5.	AAVSO.   6.  Shugarov et al. 2002.  7.	Krautter et al. 1984.  8.  Diaz et al. 1995.  9. Diaz \& Steiner 1994.   10. This work.  11. Diaz \& Steiner 1991.  12. Schaefer et al. 2010.  13. USNO-B1.0 B1 magnitude.  14. Lipkin \& Leibowitz 2008.}
\label{Table1}
\end{deluxetable}

\begin{deluxetable}{lccccccccc}
\tabletypesize{\scriptsize}
%\rotate
\tablewidth{0pc}
\tablecaption{Novae With Large $\Delta$m Values}
\tablehead{\colhead{Nova}	& \colhead{Year}	&	\colhead{Class($t_3$) (d)}	&	\colhead{$T_{th}$ (yr)}	&	\colhead{$T_{obs}$ (yr)}		& \colhead{$\Delta m$} & \colhead{$P_{orb}$ (hr)} & \colhead{Long-Lasting SSS?} & \colhead{Magn. WD?} & \colhead{Post-Decline?}	}
\startdata
V723 Cas		&  1995  &  J(299)  	&  37  &	$>$14	&	3.01		&	16.62	&	Yes	&		Yes	&	\ldots\tablenotemark{a}	\\
V1500 Cyg	&  1975  &  S(4)		&  0.1  &	\ldots	&	3.6		&	3.35		&	Yes	&		Yes	&	0.09 mag yr$^{-1}$	\\
V1974 Cyg	&  1992  &  P(43)		&  4.1  &	2		&	$>$4.12	&	1.95		&	No	&		Yes		&	\ldots\tablenotemark{a} \\
GQ Mus		&  1983  &  P(45)		&  4.4\tablenotemark{b}  &	10		&  $\approx$3.45  &	1.42		&	Yes	&		Yes	&	0.08 mag yr$^{-1}$	\\
CP Pup		&  1942  &  P(8)		&  0.3  &	\ldots	&	$>$4.96	&	1.47		&	\ldots  &		Yes	&	0.016 mag yr$^{-1}$	\\
T Pyx		&  1866  &  ?(62?)		&  7.1?  &	\ldots	&	4.7		&	1.83		&	Yes	&		Yes		&	0.026 mag yr$^{-1}$	\\
V4633 Sgr	&  1998  &  P(44)		&  4.2  &	? &	$>$2.3	&	3.01		& Ambiguous &      	Yes		&	\ldots\tablenotemark{a}	\\
RW UMi		&  1956  &  ?(140)		&  24  &	\ldots	&	$>$2.52	&	1.42		&	\ldots  &		Yes		&	0.03 mag yr$^{-1}$	\\
\enddata
\tablenotetext{a}{There is too little time since the end of this eruption tail to significantly measure any small slope in the post-eruption decline.}
\tablenotetext{b}{The theoretical formula for predicting $T_{th}$ uses the `intrinsic decay time', which is different from $t_3$ for the case of GQ Mus.  Hachisu \& Kato (2008) give the `intrinsic' $t_3$ value to be 122 days, which gives a substantially longer $T_{th}$ value.}
\label{SUM}
\end{deluxetable}

\clearpage
\begin{figure}
\epsscale{1.0}
\plottwo{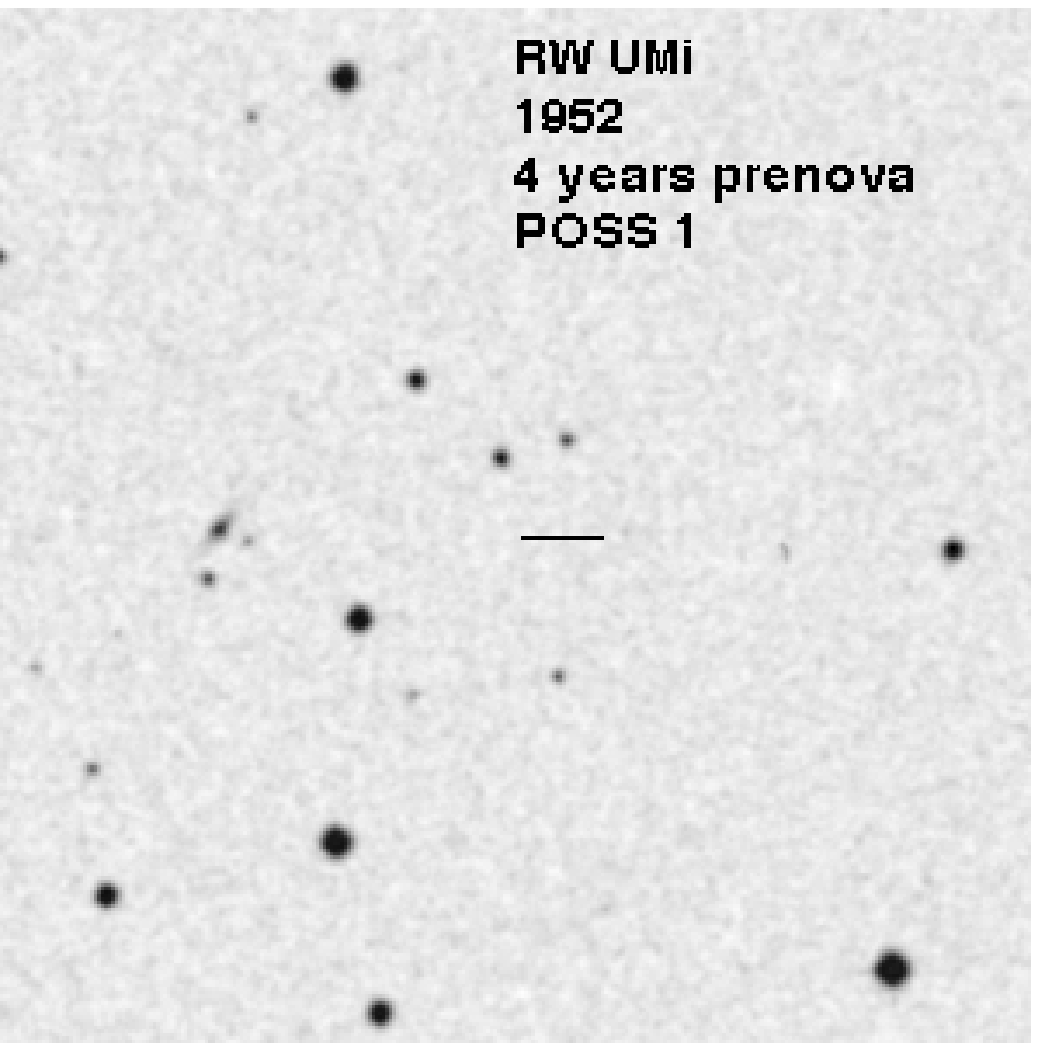}{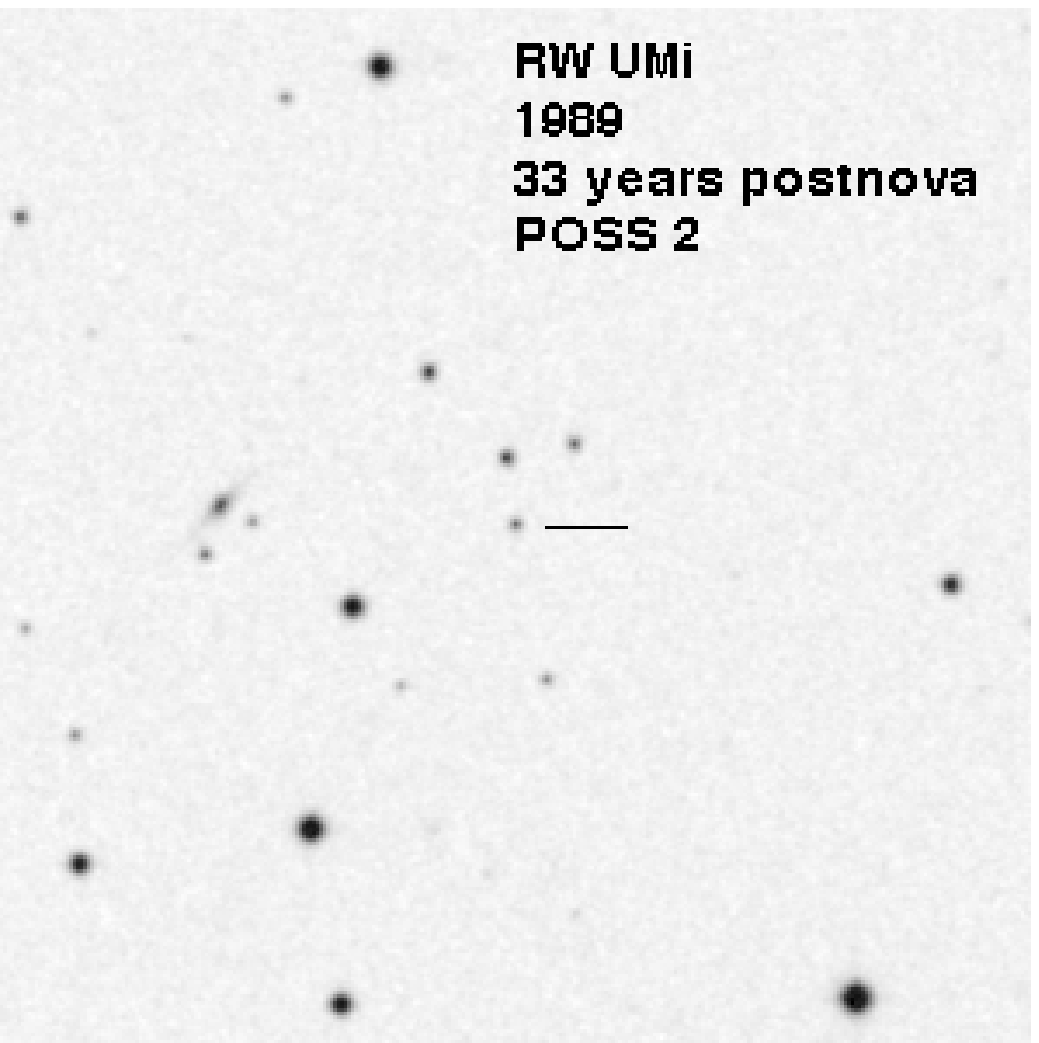}
\caption{
RW UMi before and long after eruption.  The nova erupted in 1956.  The first epoch Palomar Sky Survey plates were taken four year before the eruption and showed the system to be below the plate limits (B$>$21).  The second epoch Palomar Sky Survey plates were taken 33 years after the eruption (so the system is certainly in quiescence), and they show the system  at B=18.33.  This simple comparison demonstrates that RW UMi is much brighter in its post-eruption quiescence than in its pre-eruption quiescence.  Both panels are for the blue plates, north is up and east to the left, the fields are 5 arc-minutes square, and the position of RW UMi is just to the left of the short line.}
\end{figure}

\clearpage
\begin{figure}
\epsscale{1.0}
\plotone{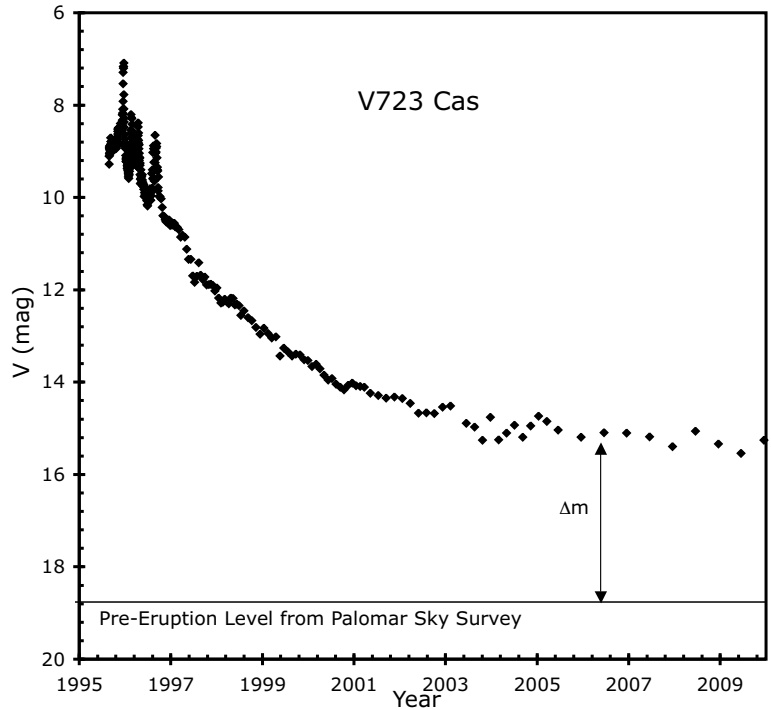}
\caption{
V723 Cas light curve 1995-2009.  The eruption light curve shows prominent large amplitude jitter, or flares, around the time of peak.  The tail of the eruption goes flat by late 2003, and this is the end of the eruption.  The pre-eruption magnitude is V=18.76 as based on the Palomar sky survey plates (Collazzi et al. 2009), with this level indicated by the horizontal line near the bottom.  The key point is that V723 Cas has its eruption ended and the nova is greatly brighter than it was prior to the eruption, which is to say that $\Delta$m is very large.  The interval 2003-2009 is too short to allow any measure of a slow post-eruption decline.  The entire light curve is based on over 20,000 visual observations from the {\it AAVSO}.}
\end{figure}

\clearpage
\begin{figure}
\epsscale{1.0}
\plotone{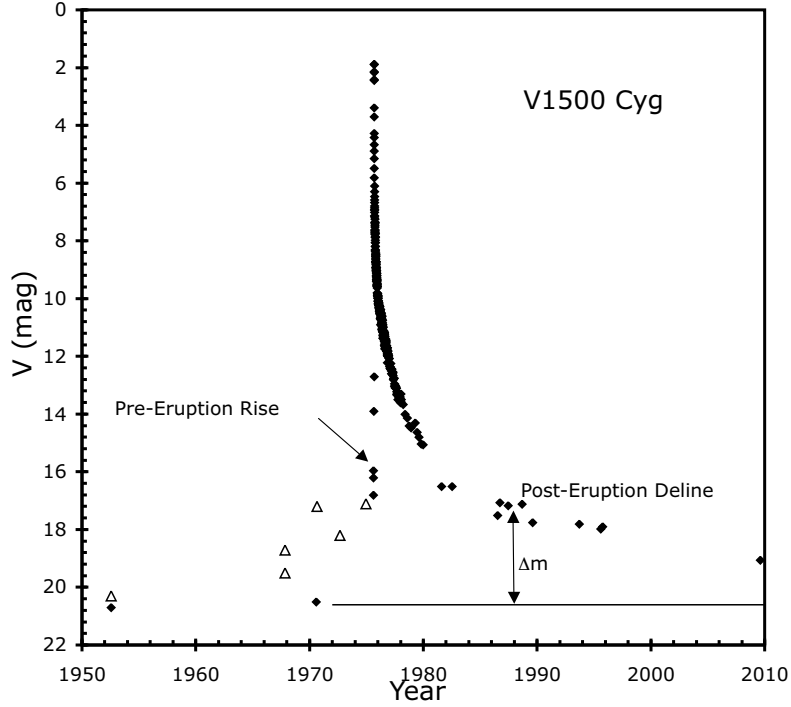}
\caption{
V1500 Cyg light curve 1952-2009.  Before the eruption (and before the pre-eruption rise), the nova was at V$\approx$20.6.  The eruption ended by 1982, as V1500 Cyg is the all-time fastest classical nova eruption and its thermonuclear runaway is certainly not extending past 1982 to 2009.  The post-eruption quiescent level is much brighter than before the eruption.  We also see a long slow post-eruption decline.  This light curve displays two hallmarks of the V1500 Cyg stars; the high value of $\Delta$m and the long slow post-eruption decline.  This light curve is constructed from binned AAVSO data, the pre-eruption data collected in Collazzi et al. (2009), post-eruption data from Pavlenko (1983; 1990), Kaluzny \& Semenuk (1987), Kaluzny \& Chlebowski (1988), Szkody (1994), Schmidt et al. (1989), DeYoung (1993), Semeniuk et al. (1995), and Somers \& Naylor (1999), while the July 2009 point is our own measured at McDonald Observatory.  Empty triangles are upper limits.  We have converted B and R magnitudes to the V-band with B-V=0.8 and V-R=0.5 (Szkody 1994).}
\end{figure}

\clearpage
\begin{figure}
\epsscale{1.0}
\plotone{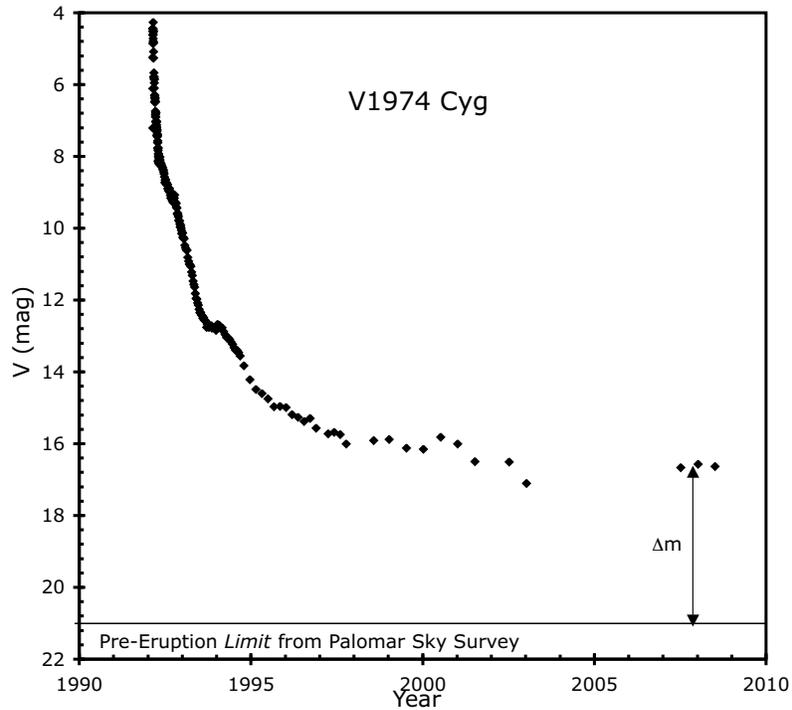}
\caption{
V1974 Cyg light curve 1992-2008.  Before the eruption, even the deep Palomar sky survey plates showed a blank field in the place of the nova, so the {\it limit} of $B>21$ (Collazzi et al. 2009) is illustrated as a line across the bottom.  The nova eruption light curve has gone flat by 2002, indicating the end of the eruption.  The quiescent level is greatly brighter than the pre-eruption level, and this shows V1974 Cyg to be a high-$\Delta$m star.  The light curve after 2002 is too short in duration to be sensitive to any small non-zero decline rate.  This light curve is constructed from 9700 visual magnitudes recorded by the {\it AAVSO}.}
\end{figure}

\clearpage
\begin{figure}
\epsscale{1.0}
\plotone{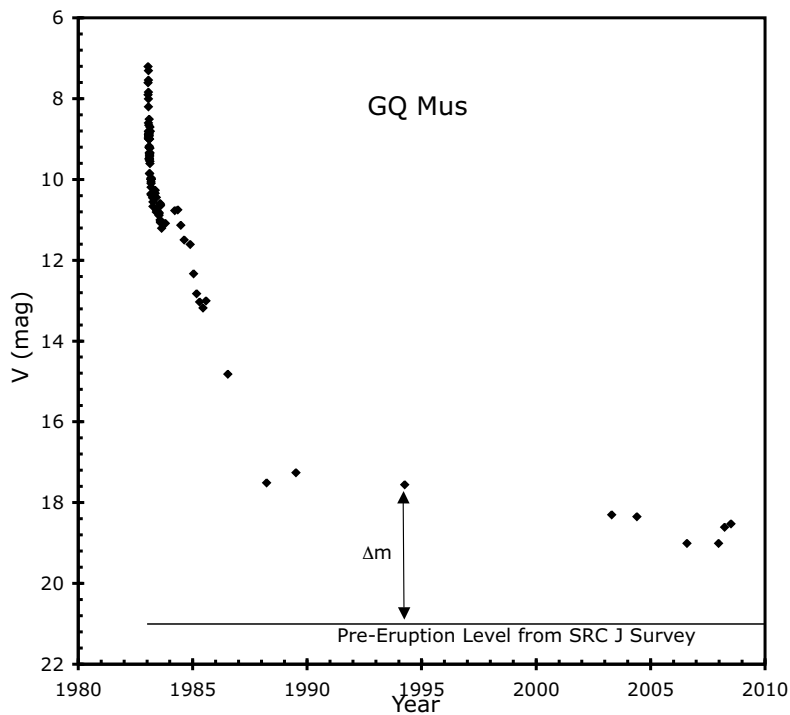}
\caption{
GQ Mus light curve 1983-2008.  Before the eruption, the pre-nova is barely visible on the SRC J plates, which gives a V-band magnitude of roughly 21 (Krautter et al. 1994), as indicated by the horizontal line across the bottom of the plot.  The light curve goes nearly flat in 1988, although there is a very slow post-eruption decline from 1988-2008.  Throughout this post-eruption interval, the nova is greatly brighter than the pre-eruption level (which is to say that $\Delta$m is large).  This light curve is constructed from prediscovery observations in Bateson (1983), the discovery observations of Liller (1983), a further IAU Circular with Nikoloff et al. (1983), 209 visual magnitudes recorded by the {\it AAVSO},  a large multicolor set of optical and infrared photometry in Whitelock et al. (1984), and late photometry of Diaz \& Steiner (1989), Diaz \& Steiner (1994), and Diaz et al. (1995).}
\end{figure}

\clearpage
\begin{figure}
\epsscale{1.0}
\plotone{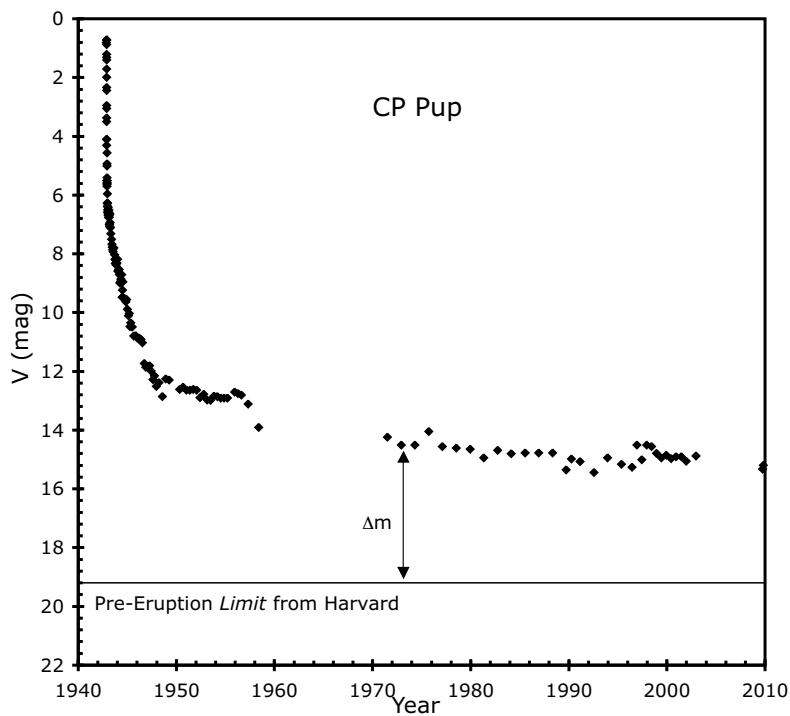}
\caption{
CP Pup light curve 1942-2009.  We have recently examined the deepest plates in the Harvard plate archives (using our modern measure of a sequence of comparison stars) and find that the pre-nova is never visible, with the deepest limit of B=19.4 (corresponding to V=19.2).  This pre-eruption {\it limit} from Harvard is indicated by a horizontal line across the bottom.  The light curve goes nearly flat sometime in the 1960's.  The light curve from 1971 to 2009 displays a small but significant steady decline.  After the eruption ended, the post-eruption light level is much brighter than the pre-eruption light level.  This light curve is constructed from 1160 visual magnitudes recorded by the {\it AAVSO}, late photometry of Diaz \& Steiner (1991), plus our two magnitudes from late 2009 with the SMARTS 1.3-m telescope on Cerro Tololo.}
\end{figure}

\clearpage
\begin{figure}
\epsscale{1.0}
\plotone{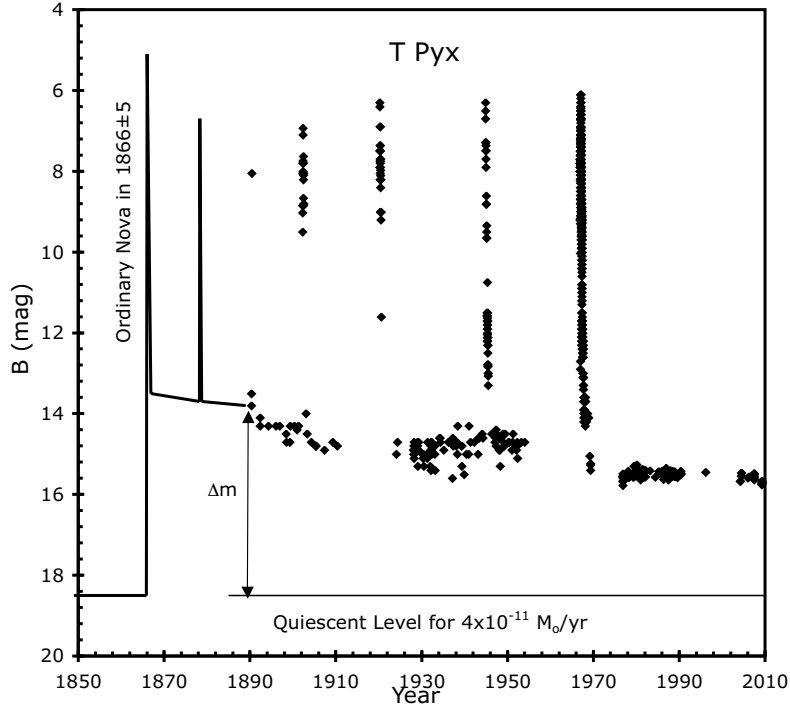}
\caption{
T Pyx light curve 1850-2009.  The high-$\Delta$m event on T Pyx was the {\it ordinary} nova event from the year 1866$\pm$5.  The existence of this event was proven by Schaefer et al. (2010) when {\it HST} images showed the nova shell to be expanding (with no deceleration) at a slow velocity (much too slow to be like the later recurrent nova events) and with a large mass (much too large to be ejected by a recurrent nova event).  To get such a large ejecta mass on a near-Chandrasekhar mass WD, theory strongly says that the accretion rate must be very small, with this being close to the accretion rate as expected to be driven by the angular momentum loses from gravitational radiation alone.  With this accretion rate, the pre-nova system brightness must have been around B=18.5 mag, as indicated by the horizontal line.  We also have observations of the quiescent level from {\it before} the 1890 eruption, with B=13.8.  Thus, the $\Delta$m value across the 1866 eruption is roughly 4.7 mag.  T Pyx displays a significant secular decline (that does not appear to be affected by the recurrent nova events) from 13.8 in 1890 to 15.7 in 2009.  The schematic light curve from 1850 to 1890 is based on the reconstruction in Schaefer et al. (2010), as discussed in Section 3.6.  The dots indicate observed data points, as presented in Schaefer (2005; 2010), Schaefer et al. (1992), Eggen et al. (1967), and Landolt (1970).}
\end{figure}

\clearpage
\begin{figure}
\epsscale{1.0}
\plotone{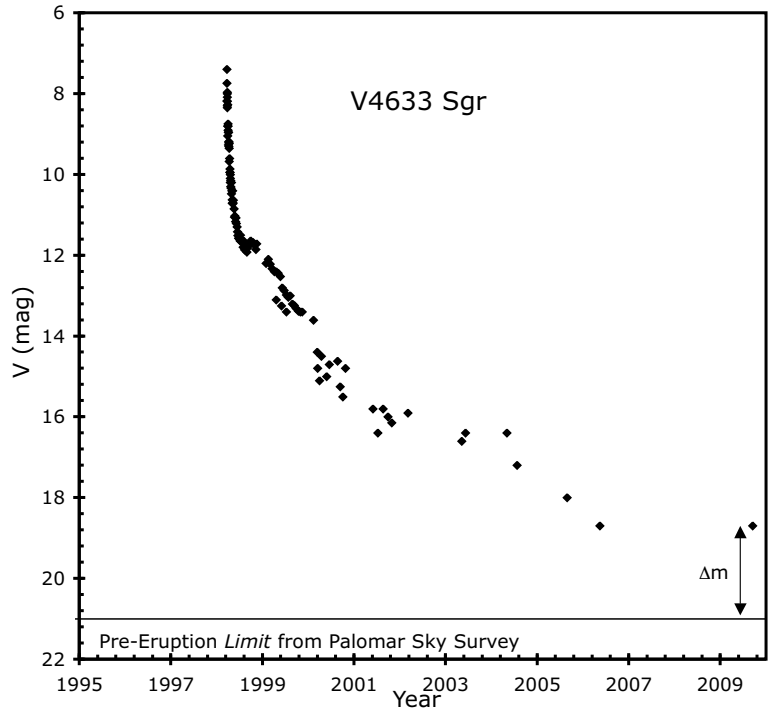}
\caption{
V4633 Sgr light curve 1998-2009.  The pre-nova was invisible on the Palomar sky survey, so the magnitude is fainter than 21 (Collazzi et al. 2009), with this {\it limit} being shown as a horizontal line across the bottom of the plot.  The light curve is constant from 2006 to 2009, with the brightness level greatly higher than the pre-eruption limit.  This light curve is constructed from 772 visual magnitudes recorded by the {\it AAVSO}, multicolor photometry of Lipkin \& Leibowitz (2008), plus our magnitude from the summer of 2009 with the 0.8-m telescope at McDonald Observatory.}
\end{figure}

\clearpage
\begin{figure}
\epsscale{1.0}
\plotone{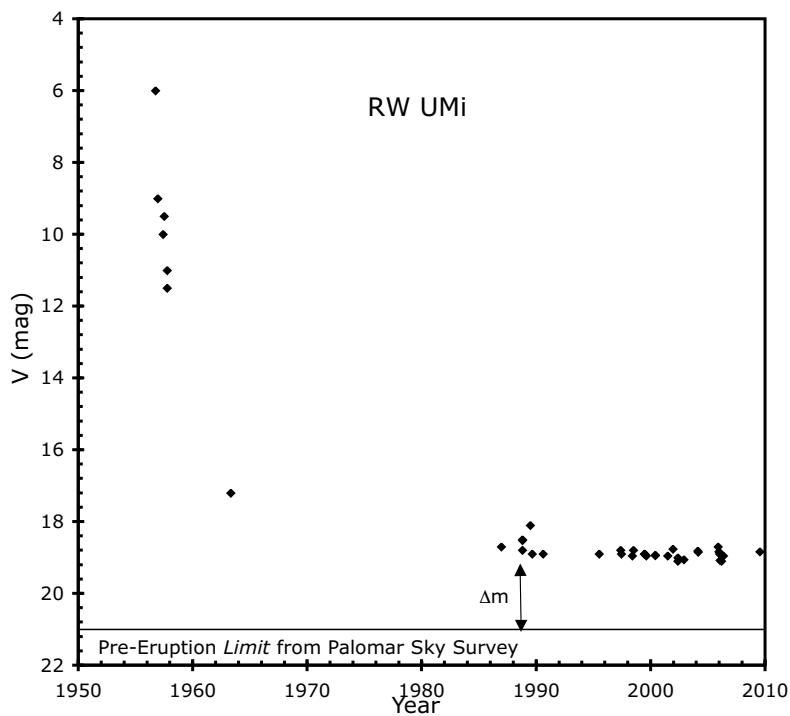}
\caption{
RW UMi light curve 1956-2009.  This nova was only discovered in 1963, based on archival photographs, so the eruption light curve is sparse.  The pre-nova is invisible on the Palomar sky survey, so B$>$21 (Collazzi et al. 2009).  The light curve long after the eruption is over is greatly brighter than this pre-eruption limit.  Detailed work by Bianchini et al. (2003) and Tamburini et al. (2007) shows that the post-eruption light curve is slowly declining.  This light curve is constructed from archival data during the eruption as given in Kukarkin (1963) and Ahnert (1963), plus the many late-time observations compiled and observed by Bianchini et al. (2003) and Tamburini et al. (2007), plus our magnitude from the summer of 2009 with the 0.8-m telescope at McDonald Observatory.}
\end{figure}

\clearpage
\begin{figure}
\epsscale{1.0}
\plotone{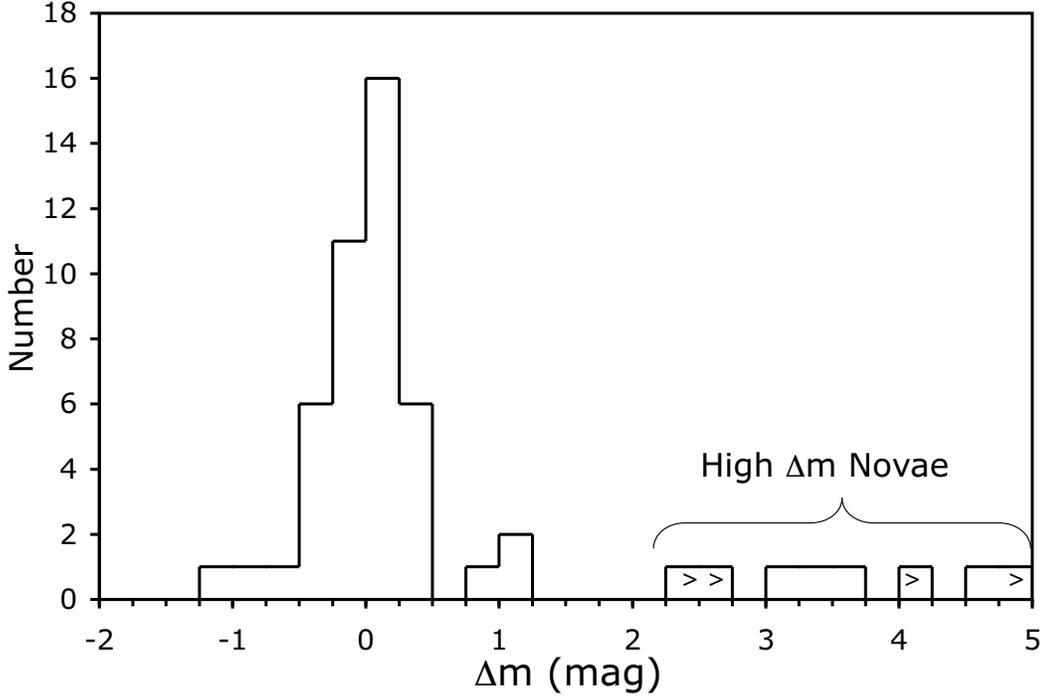}
\caption{
Distribution of $\Delta$m.  This histogram of the brightness change across nova eruptions ($\Delta$m=$m_{pre}-m_{post}$) is from measurements in Collazzi et al. (2009) and this paper.  The `$>$' symbol indicates a lower limit on $\Delta$m.  We see that most novae have values around zero, with the expected scatter due to ordinary variations in quiescence, which indicates that most nova eruptions do not significantly change the long-term accretion rate in the systems.  However, we also see that eight of the nova have significantly large $\Delta$m, with all or most being more than a factor of ten times brighter (i.e., $\Delta$m$>$2.5 mag) after eruption than before eruption.  These changes are all much too large to be due to measurement error, ordinary fluctuations, or flickering.  With these systems' light being dominated by their accretion light, something related to the nova eruption must have made their accretion rates increase by over a factor of ten.  We are naming this group of high-$\Delta$m stars after their prototype V1500 Cyg.}
\end{figure}

\end{document}